\documentclass[conference]{IEEEtran}
\IEEEoverridecommandlockouts
\usepackage{cite}
\usepackage{amsmath,amssymb,amsfonts}
\usepackage{algorithmic}
\usepackage{graphicx}
\usepackage{textcomp}
\usepackage{xcolor}

\usepackage{hyperref}
\usepackage{mathtools}
\usepackage{multirow}
\usepackage{listings}
\usepackage{booktabs} 
\usepackage{float}
\usepackage{array}

\def\BibTeX{{\rm B\kern-.05em{\sc i\kern-.025em b}\kern-.08em
    T\kern-.1667em\lower.7ex\hbox{E}\kern-.125emX}}
\begin{document}

\title{Smart Contract Coordinated Privacy Preserving Crowd-Sensing Campaigns}

\author{
\IEEEauthorblockN{
    Luca Bedogni\IEEEauthorrefmark{1},
    Stefano Ferretti\IEEEauthorrefmark{2}
}
\IEEEauthorblockA{\IEEEauthorrefmark{1}Department of Computer Science and Mathematics, University of Modena-Reggio Emilia, Italy}
\IEEEauthorblockA{\IEEEauthorrefmark{2}Department of Computer Science and Engineering, University of Bologna, Italy}
{\tt\small luca.bedogni@unimore.it, s.ferretti@unibo.it}
}

\maketitle

\begin{abstract}
Crowd-sensing has emerged as a powerful data retrieval model, enabling diverse applications by leveraging active user participation. However, data availability and privacy concerns pose significant challenges. Traditional methods like data encryption and anonymization, while essential, may not fully address these issues. For instance, in sparsely populated areas, anonymized data can still be traced back to individual users. Additionally, the volume of data generated by users can reveal their identities. To develop credible crowd-sensing systems, data must be anonymized, aggregated and separated into uniformly sized chunks. Furthermore, decentralizing the data management process, rather than relying on a single server, can enhance security and trust. This paper proposes a system utilizing smart contracts and blockchain technologies to manage crowd-sensing campaigns. The smart contract handles user subscriptions, data encryption, and decentralized storage, creating a secure data marketplace. Incentive policies within the smart contract encourage user participation and data diversity. Simulation results confirm the system's viability, highlighting the importance of user participation for data credibility and the impact of geographical data scarcity on rewards. This approach aims to balance data origin and reduce cheating risks.
\end{abstract}


\section{Introduction}
\label{sec:introduction}
Crowd-sensing has long been recognized as a data retrieval model that can facilitate a plethora of diverse applications. It opens the doors to service development not only by tech giants, which typically hold big data, but also by smaller developers who lack in-house data but often have innovative service ideas. These developers leverage active user participation, turning users into both data generators and service consumers.

However, a common challenge in crowdsensing applications is data availability. While everyone desires data, sharing it can be problematic. Data disclosure raises concerns about revealing sensitive information. A typical approach to address this issue is data encryption and anonymization \cite{KIM2022103315,KHAN2019456}. While essential in any crowdsensing campaign, it may not always be sufficient.
Let's illustrate the problem with an example. Imagine Alice lives in a sparsely populated rural area. She participates in an app that requires geo-located data. Even if the app is popular, there will likely be few participants generating data from that specific area. Consequently, it becomes easy to infer that the sparsely populated area's data was likely generated by Alice.
This simple example illustrates how the concept of data anonymization is not solely tied to the process of its encoding, but also to how such data can be aggregated with other data, thereby making it less unique in relation to the rest of the dataset.

Another aspect through which it is possible to infer information about the individual who has produced the data is the volume of the data itself. Let's imagine that Alice and Bob contribute to data production in the same area with a sensor placed in their cars. However, Alice rarely uses her car and therefore is able to produce a limited amount of data over time, while Bob commutes every day, consequently, his sensor produces a large amount of data. If the two datasets were treated as separate datasets, even though they are anonymized, it becomes easy to trace back to who produced them, once a bit of information about the two users is available.

These examples highlight some fundamental aspects for the development of credible and trusted crowd-sensing systems. First, the data must be anonymized. Second, the data should only be made available if there is a certain number of participating users that allow the creation of an 'anonymization set', making it possible to understand that Alice has produced some data that are in a certain dataset, but the number of participants is such that it becomes difficult to trace back to the data itself. Third, the data must be separated into uniformly sized chunks, in order to decouple the size of the dataset from the user who produced it. Fourth, the mechanism through which the crowd-sensing campaign is carried out should be decentralized and not tied to the presence of a single server managing all the information \cite{chen2021blockchain,gigli2024decentralization}. The presence of a server, although it makes the implementation of a crowd-sensing system easier, presents some problems. The server is typically the single point of failure of the system. Moreover, the server becomes a trusted third party, which cannot always be trusted. Finally, the presence of a single system that maintains all the information makes the system prone to attacks. If an intruder manages to enter the single node, they can obtain all the information.

In this paper, we describe a system for managing crowd-sensing campaigns based on smart contracts and blockchain technologies. The campaign is entirely managed by a smart contract (SC), which collects user subscriptions, provides the keys with which to encrypt the data that will be uploaded to a decentralized storage, e.g., the InterPlanetary File System (IPFS) \cite{10.1145/3544216.3544232}. Once a sufficient number of subscriptions have been received, and once the data provided by the users have been verified, the SC reveals to the subscribers the locations from which to download the data and the keys to decrypt them. The SC thus represents a data marketplace for crowd-sensing. Furthermore, the SC implements incentive policies based on rewards, in order to favor those who cooperate within the system and penalize those who instead try to take advantage of the data without contributing in some way.

We confirm the viability of the proposed system through simulation.
The obtained results show that user participation plays a fundamental role in the success of a crowdsensing campaign. The greater the participation, the better the data credibility. Additionally, the results highlight an important observation: as the number of users in a specific location increases, their associated rewards decrease. This phenomenon indicates that data scarcity from a particular geographical area makes such data more desirable hence more highly compensated. While this limits the possibility of paying more for false data compared to genuine data from heavily frequented locations, we also expect this phenomenon to encourage users to contribute data from diverse places, balancing the geographical origin of the data and reducing the risk of cheating.

The remainder of this paper is organized as follows. 
Section \ref{sec:back} provides the background needed for the rest of the paper, as well as the state of the art. 
Section \ref{sec:archi} describes the proposed system architecture. 
Section \ref{sec:perf} describes the methodology to assess the proposed system, and presents the experimental results. 
Section \ref{sec:conc} provides some concluding remarks.

\section{Background}
\label{sec:back}
There is an extensive body of literature on architectures for crowdsensing and a significant amount of research on crowdsensing systems that leverage blockchain technologies for traceability or participant rewarding. The novelty of this work lies in the definition of a system based on smart contracts that not only encourages participation but also minimizes the possibility of de-anonymization of participants as much as possible. We believe that this mechanism can have interesting applications across all contexts of geolocated data production.
In rest of this section, we review the state of the art for the different areas pertaining to this work.

\subsection{Crowdsensing Applications}
Crowdsensing is a set of techniques to gather sensed data without the need to deploy a dedicated sensor infrastructure. However, data coming from participants should be properly treated, in order to safeguarding the personal data of contributors on one hand, and checking data quality, on the other.

Privacy concerns primarily have to deal with revolve around the protection of contributors' activities. The main example is users' movement patterns and routines, as geolocated data can reveal personal habits and places of interest \cite{Bedogni2018UserPrivacy}. In the literature, some works present strategies to maintain contributor anonymity \cite{Cheng2022Privacy,KIM2022103315}. Recent advancements also explore protocols that allow for secure communication between contributors and servers without excessive data exposure, using correlation metrics to balance privacy and functionality \cite{montori2023privacy}. It is also important to consider how such data can be transferred to the campaign manager \cite{ALOI201774}.

As already mentioned, crowd-sensing relies to the rewarding strategies to incentivize users to participate. However, the reward system is intrinsically linked to privacy; platforms must identify contributors to compensate them appropriately \cite{capponi2019survey,Xu2022Incentives}. Thus, mechanisms to rewarding contributors while preserving their anonymity may be used, where a feedback-based protocol enables contributors to control their level of disclosed information \cite{BEDOGNI2023103634}.

\subsection{Blockchain-based Crowd-sensing}
In the literature, there is a good state of the art on papers that merge the ideas of crowd-sensing and distributed ledgers \cite{chen2021blockchain,gigli2024decentralization,zou2019crowdblps}. Quite often, however, the blockchain is merely used as a data exchange system for ensuring non-repudiation and non-tampering, without complete architectural integration and without resorting to smart contracts as a general coordinator for the whole crowd-sensing campaign. 

More recent studies propose their own decentralized ledger technologies to accommodate the crowd-sensing model. For example, the study in \cite{huang2022blocksense} introduces BlockSense, a blockchain for crowd-sensing that is constructed on a new consensus algorithm named Proof-of-Data (PoD). This algorithm enables miners to verify data quality instead of solving pointless puzzles. To accomplish this, the authors employ zk-SNARKS and Homomorphic Encryption on each data point. Another study \cite{an2020lightweight} suggests a similar model based on Delegated Proof of Reputation (DPoR) as the consensus mechanism and multiple contracts to regulate the actors' behaviour. The aspect of fair rewarding is considered in ABCrowd \cite{kadadha2020abcrowd}, where the authors introduce the well-known reverse auction paradigm into a decentralized ecosystem as an incentive. They also design a crowd-sensing system to counteract misbehaving workers over Ethereum \cite{kadadha2020sensechain} by allowing actors to interact with each other through smart contracts. However, this solution is hardly sustainable due to Ethereum's high transaction costs.

On a larger scale, there are works that focused on the creation of data marketplaces \cite{zichichi2020framework}, but without going into specific crowdsensing aspects \cite{bonini2023proof}.
A previous work by the authors of this paper proposed a decentralized system for mobile crowd-sensing \cite{gigli2024decentralization}. With respect to this paper, in that work the smart contracts were basically used for rewarding the activities of participants and for storing the position of the data. In this work, we focus more on the aspects related to reaching a certain threshold of participant, in order to increase the anonymity set necessary to ensure privacy.

\section{System Architecture and Protocol}
\label{sec:archi}
In this section we describe our system model and the scenario in which we study it. Specifically, we target crowdsensing campaigns in which there is a number of users, also called participants, which can provide data of interest to the campaign owner through either smartphones, wearables or in generale mobile devices \cite{Fortino2023Wearable}. While in highly populated areas this behavior hinders the detection of participants by a malicious entity, in low populated areas it is easier, as less users are contributing to the campaign, hence making them easier to be recognized. To overcome this limitation, we leverage blockchain technology and in particular Smart Contracts, to collect data only if privacy requirements are met. 

\begin{figure}
    \centering
    \includegraphics[width=.7\linewidth]{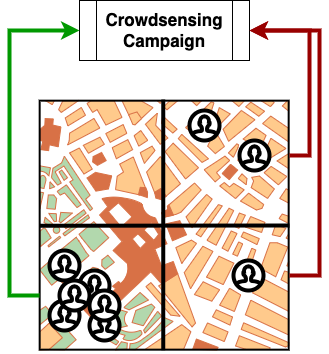}
    \caption{Scenario of our study. Users in an area may report data to the campaign, only if a minimum population is available due to privacy concerns.}
    \label{fig:scenario}
\end{figure}

In Figure \ref{fig:scenario} we depict the scenario of our study. In this case, if users in the right side of the map report data to the crowdsensing campaign they may be identified by an attacker, as there are less other users in the area. Conversely, users on the bottom left area can report it with higher privacy guarantees, as other users in the same area provides better protection against attacks.  

\subsection{Actors}
Our system is composed of the following actors:
\begin{itemize}
\item A set of data sources (hereafter "Users") willing to provide their data for in exchange for a reward.
\item A Smart Contract (SC) tasked with coordinating the data collection campaign and rewarding those who have contributed data.
\item Verifiers who are in charge of verifying that the data provided by a user is credible.
\item A Decentralized File Storage (DFS) where encrypted data is store, such as IPFS or a (sovereign) cloud storage.
\end{itemize}

\subsection{Protocol}
\begin{figure*}[h]
    \centering
    \includegraphics[width=.6\linewidth,height=280px]{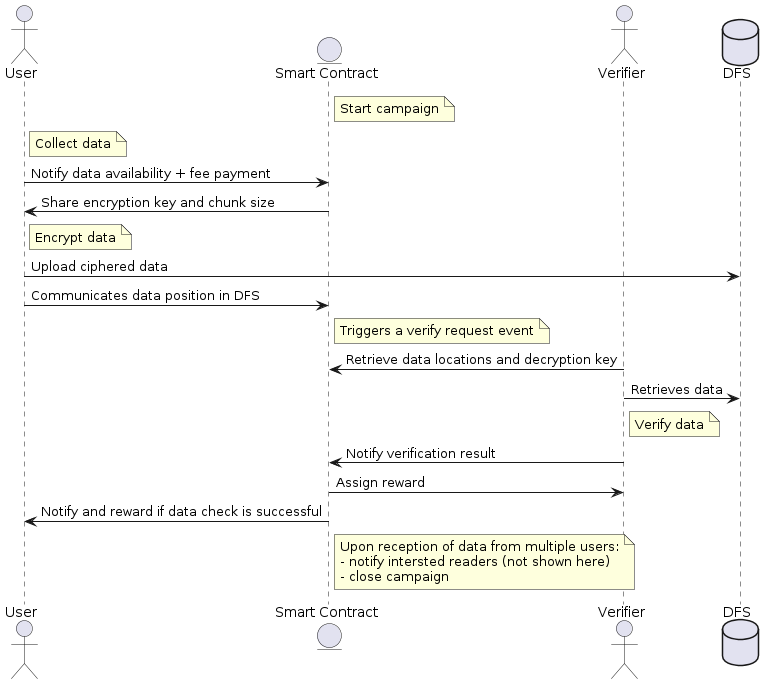}
    \caption{System protocol}\label{fig:prot}
\end{figure*}

The protocol is shown in Figure \ref{fig:prot} and is as described in the following. For the sake of a simpler presentation, only one user is considered in the description, but needless to say, multiple users should act as data sources. The same holds for verifier, who may actually be more than one.

\begin{enumerate}
\item The SC is created and the crowd-sourcing campaign is open. This operation is in charge to the person (or entity) who is interested in the data gathering from multiple sources. Following this operation, we envision that there will be a data collection promotion campaign aimed at incentivize users to participate. However, this activity may involve several engagement actions not accounted for in the protocol, as they are independent of it.
\item The interested User collects the data to be uploaded to the DFS.
\item When the data source is ready, it notifies the SC and pays a fee. This fee serves as an incentive to discourage cheating and prevent denial-of-service (DoS) attacks.
\item The SC shares an encryption key with which to cipher the data and the size of the data chunks to be used to partition the data. The idea of using equal-sized chunks prevents outsiders from inferring the size of the data produced by a single user.
\item The User encrypts the data and uploads them to the DFS, as separate data chunks.
\item The User communicates to the SC the position of the data in the DFS.
\item The SC triggers a "verification request" event so as to wake up the interested Verifiers that listen for novel data checks to accomplish.
\item The Verifiers will ask the SC for the data locations and the key to decrypt data. Multiple requests can be received by different Verifiers. In this work, we assume that just one requestor will be elected as Verifier for a given dataset provided by one user. For security reasons, it might be possible to ask for redundant verifications that would add more trust on the verification results, in the case Verifiers cannot be completely trusted or if they can fail. However, we do not consider this aspect here.
\item The Verifier retrieves the data from the DFS.
\item The Verifier checks the data and notifies the SC if everything is okay, i.e., data are encoded using the proper encoding format, appear to be valid, etc.
\item (Assuming the result of this verification is trustworthy,) the SC pays the Verifier.
\item If the data passed the verification, the User is rewarded. Conversely, users whose data did not pass the Verifier's test are not rewarded (and therefore lose the fee).
\item When the number of Users who have passed this check is above the minimum participation threshold, the SC notifies the interested parties, passing the links from the various data chunks, and passing the key to decrypt them, and the campaign is closed.
\end{enumerate}

It is worth mentioning that in this description and in the figure, we focused primarily on the data retrieval. Thus, we did not focus on the aspects related to the subscription of potential data readers that pay a fee in order to obtain the data. We claim this aspect is straightforward, i.e., potential readers ask for the data by paying a fee that in turn receive data locations of the key to decrypt. For the sake of simplicity, Figure \ref{fig:prot} does not show this part of the system protocol.

\section{Performance Evaluation}
\label{sec:perf}
In this section we describe our simulator and we present the performance evaluation of our proposed system.

\subsection{Simulation Details}
We have built a custom made simulator, implemented in Python, to simulate the movement of users in an fixed area. 
The simulated area was a non-toroidal two dimensional space. For human movement and simulation advancements, we used a classic time-stepped simulation approach, in which users move from one point to another. This approach has already been followed in literature, to test the system under a varying number of users \cite{Fortino2017Simulation}
In the simulated area, several smart contracts are defined over smaller areas, in which data is requested to users. We have varied both the number of participants in the area, to study the effects of different population densities, as well as the minimum number of persons required by each SC. Users move freely in the area, with a classic random walk movement type. Whenever they are in the area of a SC, they compete with other users to send data. 

We have implemented a system where the SC selects the minimum number of users to satisfy its needs, matching the privacy requirements of each user. Upon selecting users, the SC also pays to each of them the corresponding reward.
If a user is selected, it raises the reward asked for the next iteration, while if it is not selected it decreases it. This balances the trade-off between the offer of data from users and the real demand from any SC. In our simulation, we have set the initial rewards asked by users to 1, and they increment or decrement it as explained by 0.1. We have run multiple simulations of a full day of movements.

\subsection{Simulation Analysis and Results}
\begin{figure}
    \centering
    \includegraphics[width=\linewidth]{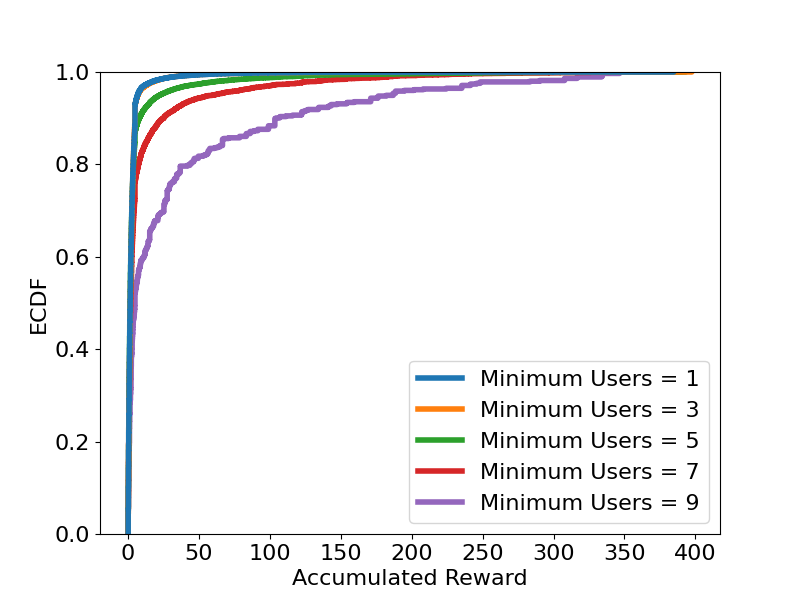}
    \caption{Accumulated reward ECDF varying the minimum number of persons for each SC. }
    \label{fig:reward-ecdf}
\end{figure}
The first analysis we perform is shown in Figure \ref{fig:reward-ecdf}, and it is the Empirical Cumulative Distributive Function (ECDF) on the accumulated reward by each user, varying the minimum number of users providing data simultaneously. As it can be seen, when the minimum number of persons in a SC area is low, users do not accumulate considerable rewards. This is due to the fact that when a low number of persons is required to fullfil the SC needs, several users may satisfy it, hence the offer of data is much greater than its demand, hence the overall average reward decreases.

\begin{figure}
    \centering
    \includegraphics[width=\linewidth]{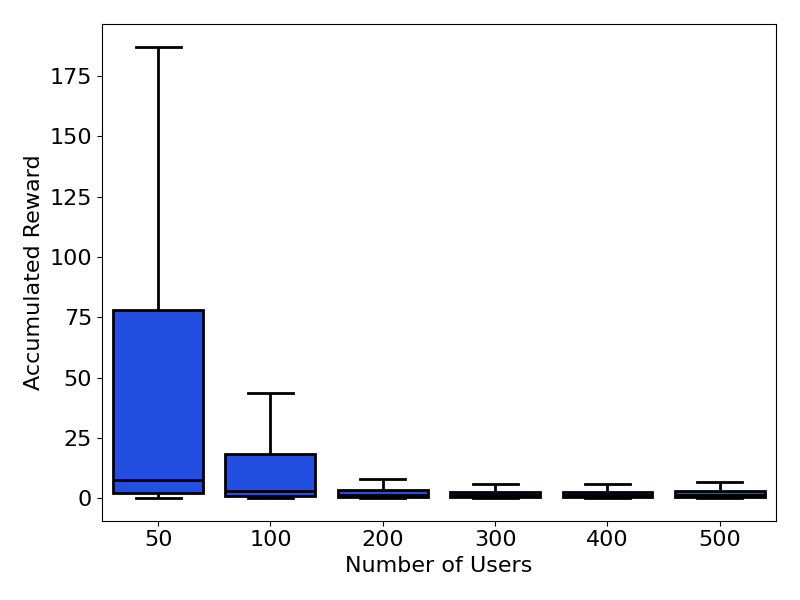}
    \caption{Accumulated reward versus the population density. A higher number of people translates into a higher data offer, hence reduced rewards.}
    \label{fig:reward-vs-nhumans}
    \vspace{-0.3cm}
\end{figure}
The second analysis we show is also on the accumulated reward by users, versus the population density. As it can be seen, when the number of users is low, the overall reward is higher, as there are fewer users that can provide the data to the SC. As the number of users increases, the number of people that can offer data is higher, hence the data of the single human being is less valuable, eventually leading to lower rewards. Although the median values are similar, when the population density is low there are some users which succeed in gaining significantly higher rewards.

\begin{figure}
    \centering
    \includegraphics[width=\linewidth]{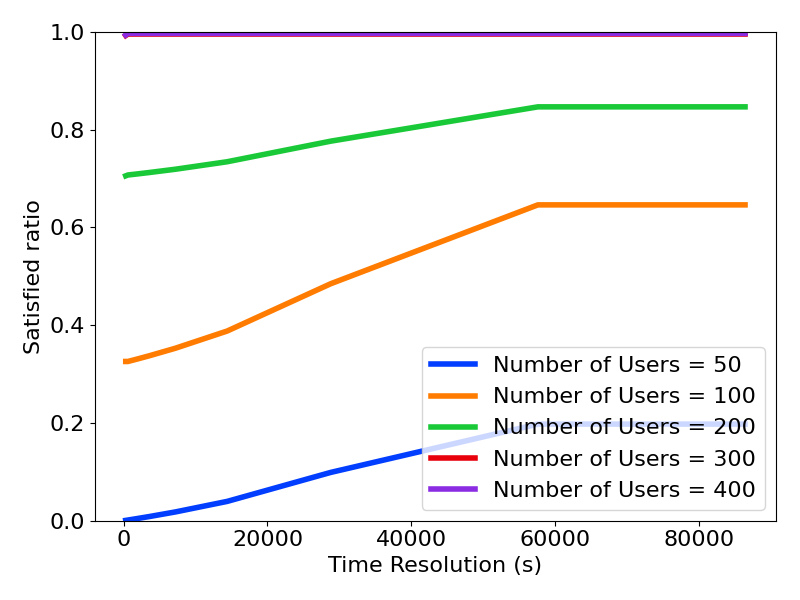}
    \caption{Satisfied ratio versus data freshness. Decreasing the requirements for data freshness allows SC to be satisfied with a lower population density. }
    \label{fig:satisfied-vs-freshness}
    \vspace{-0.3cm}

\end{figure}
In Figure \ref{fig:satisfied-vs-freshness} we show the ratio of satisfied SCs, versus the data freshness. The data freshness is the maximum time frame in which data obtained from participants is considered valid. A higher time frame means that the data can be considered valid for a longer period of time, while a lower time frame requires more frequent updates to the SC. This scenario happens when there is heterogenous data required by different SC, which changes at a different pace considering the kind of data itself. In Figure \ref{fig:satisfied-vs-freshness} it is shown the key importance this parameter has. If we look at the case with a low population density, increasing the data freshness to several hours raises the ratio of satisfied SCs. This is not required for higher population densities, which can always provide recent data regardless of the freshness required by the SC. This data has been obtained considering 7 as the minimum number of people for the SC.

\begin{figure}
    \centering
    \includegraphics[width=\linewidth]{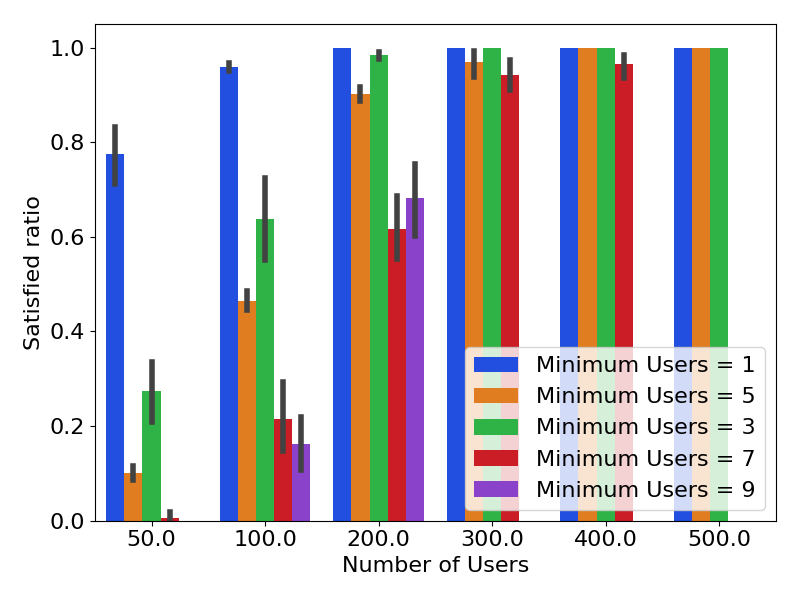}
    \caption{Satisfied ratio versus the population density. A higher number of persons needed for the SC to be satisfied requires a higher population density.}
    \label{fig:satisfied-vs-numhumans}
    \vspace{-0.3cm}

\end{figure}
Figure \ref{fig:satisfied-vs-numhumans} shows instead the satisfied ratio versus the population density, considering different requirements for the SC in terms of minimum persons. It is clear again how the population density plays a key role, as lower densities struggle to meet the needs of the SC, unless a low number of participants is requested. Instead, higher population densities can support even more tight requirements for the SC on the minimum number of persons.

\begin{figure}
    \centering
    \includegraphics[width=\linewidth]{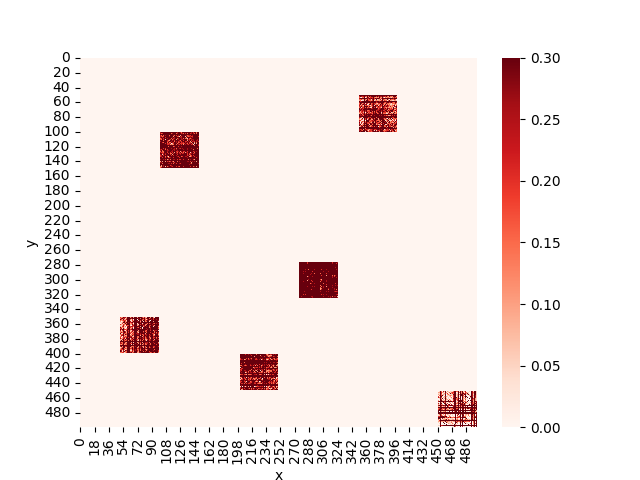}
    \caption{Heatmap of a sample run of simulation. It is evident how central smart contracts are more visited than others, hence provide higher rewards. }
    \label{fig:heatmap}
\end{figure}
Finally in Figure \ref{fig:heatmap} we show a sample run of our simulated scenario, in which we plot the average reward obtained in the different SC. Central SC, which refer to more visited areas, have a more homogeneous reward distribution, while peripheral SC tend to provide a lower reward, mainly due to less people visiting them, hence more difficulty in satisfying the SC requirements. This plot is shown considering again 7 as the minimum number of persons for any SC. 

\section{Conclusions}
\label{sec:conc}
In this paper, we presented a novel system for managing crowd-sensing campaigns using smart contracts and blockchain technologies. Our approach goes into the direction to address key challenges in crowd-sensing, such as data privacy, security, and user incentivization. Through smart contract coordination, data is encrypted, stored in a decentralized manner, and made accessible to authorized users only. The system also implements incentive policies to encourage user participation and data diversity.

Simulation results confirm the viability of our proposalcome topi. We observed that user participation plays a key role in the success of crowd-sensing campaigns. Additionally, our results highlight the impact of geographical data scarcity on rewards, suggesting that data from less frequented areas is more highly compensated. This mechanism helps balance the geographical origin of the data and reduces the risk of cheating.

There are several areas for improvement. First, we plan to explore more sophisticated anonymization techniques to further protect user privacy. Second, it could be interesting to integrate machine learning algorithms to detect and mitigate fraudulent data submissions. Finally, we plan to conduct real-world experiments to validate the system's performance.

\section*{Acknowledgements}
This work is partially supported by the European Union - NextGenerationEU within the framework of PNRR  Mission 4 - Component 2 - Investment 1.1 under the Italian Ministry of University and Research (MUR) programme ``PRIN 2022'' - grant number 2022N2NH42 SmartShires – CUP: H53D23003570006. 


\bibliographystyle{IEEEtran}

\end{document}